\documentclass{article}

\usepackage{amsmath,graphicx}
\usepackage{xcolor}
\usepackage{amssymb}
\usepackage{mathrsfs}
\usepackage{hyperref}

\usepackage[font=small,skip=2pt]{caption}
\usepackage{placeins}

\title{TOWARDS SEMANTIC RECONSTRUCTION OF INDIVIDUAL WORDS FROM fNIRS USING CLIP LOSS}

\author{S. Posso-Murillo$^{\dagger}$, N. Palladino$^{\star}$, B. Pyykkonen$^{\star}$, DY Han$^{\ddagger}$,\\ LG Sanchez-Giraldo$^{\dagger}$, and J. Bae$^{\dagger}$
\footnote{$^{\dagger}$Department of Electrical and Computer Engineering, University of Kentucky. $^{\star}$Department of Psychology, Wheaton College. $^{\ddagger}$Departments of Neurology, Neurosurgery, and Physical Medicine \& Rehab., University of Kentucky}}

\begin{document}

\maketitle

\begin{abstract}
Semantic reconstruction maps neural activity to a
word-embedding space, recovering the meaning of a perceived word instead of
selecting it from a fixed vocabulary. Functional near-infrared spectroscopy (fNIRS) carries semantic information suitable for this mapping. However, most fNIRS decoders are trained with a
squared-error objective that fits each word independently and ignores the
geometry of the embedding space. To address this limitation, we evaluate a contrastive loss based on the contrastive-language-image-pretraining (CLIP) loss, as an alternative to
mean-squared-error (MSE) for reconstructing perceived words from
fNIRS. We compare the two objectives by training a bidirectional long short-term
memory (Bi-LSTM) decoder to map fNIRS signals to word embeddings. We use GloVe-50 and T5 word embeddings as targets, across three fNIRS datasets recorded under a shared paradigm pairing each
word image with its spoken name. Performance is measured with a pairwise
matching score and open-vocabulary top-$k$ retrieval. The Bi-LSTM trained
with CLIP is the most consistent decoder across experiments. T5 produces
higher matching scores, whereas every significant retrieval result uses
GloVe-50. These results support the use of contrastive objectives as a promising direction for fNIRS
semantic decoding and motivate validation on larger datasets.
\end{abstract}


\section{Introduction}\label{sec:intro}

Artificial intelligence is redefining brain-machine interfaces (BMIs) aimed
at restoring communication for people with severe speech impairments. Using non-invasive neuroimaging, machine learning
models have used electroencephalography (EEG) and magnetoencephalography
(MEG) signals to decode typed characters~\cite{levy2025brain} or read and
heard words and sentences~\cite{d2025towards,defossez2023decoding}, and have
reconstructed perceived speech from functional magnetic resonance imaging
(fMRI)~\cite{tang2023semantic,ye2025generative}. While fMRI offers
excellent spatial resolution, it is constrained by the scanner environment.
fNIRS is a portable, motion-tolerant alternative suited for clinical
use~\cite{yucel2025fnirs}. Like fMRI, it measures the hemodynamic response to
neural activity, but through the differential absorption of near-infrared
light by oxygenated (HbO) and deoxygenated (HbR) hemoglobin.

Current fNIRS studies show that the signal carries information useful for
decoding perceived speech. \cite{zinszer2018decoding} used representational similarity analysis to demonstrate that semantic information elicited by words is
encoded in fNIRS. On the classification side, \cite{hons2026hybrid} decoded
four phonemes from a hybrid EEG-fNIRS setup over imagined and perceived
speech, \cite{cooney2021bimodal} classified overt and imagined words from a
similar hybrid setup, and \cite{rybavr2021decoding} classified imagined
words from two semantic categories using fNIRS alone.

Because these classifiers predict only words seen in training, adding a
word requires retraining with a new output class. Word embeddings avoid
this problem by representing each word as a vector in a continuous space where
semantically related words lie close together, so a decoder can predict a
vector and name the word by its nearest embedding without a fixed set of
classes. \cite{cao2021brain} mapped fNIRS into the GloVe
space~\cite{pennington2014glove} with ridge regression, and
\cite{posso2025semantic} replaced ridge with recurrent networks, removing
the dependence on hand-crafted features.

Both~\cite{cao2021brain} and~\cite{posso2025semantic} regress onto word
embeddings under a squared-error objective. MSE only requires
each prediction to stay close to its own target, with no term coupling it
to the embeddings of other semantically similar words. Therefore, the structure of
the embedding space is ignored. Additionally, in the recurrent model
of~\cite{posso2025semantic}, a regularization term that spreads the outputs of the neural network was necessary to avoid predictions to collapse.

The contrastive CLIP objective~\cite{radford2021learning}, designed to
align images and text by pulling matched pairs together and pushing
mismatched pairs apart, has become a common tool for aligning brain
activity to pretrained representations. It has been used to align fMRI with image
embeddings~\cite{scotti2023reconstructing}, MEG and EEG with self-supervised
speech audio embeddings, with substantial gains over MSE~\cite{defossez2023decoding},
and used to align word embeddings to EEg and MEG signals~\cite{d2025towards}. To our knowledge, CLIP has not been evaluated for fNIRS semantic decoding.

Based on these observations, we evaluate the CLIP objective as an alternative
to MSE for fNIRS semantic decoding. We also compare two decoding targets:
GloVe-50 word vectors~\cite{pennington2014glove}, learned from word
co-occurrence statistics, and embeddings from the pretrained transformer
language model T5~\cite{raffel2020exploring}. Language model representations
map onto fMRI  responses to language~\cite{caucheteux2022brains} and
have been used directly as decoding targets for EEG and
MEG~\cite{d2025towards}.\footnote{Code and data to replicate the results
will be made available upon publication.}

\section{METHODOLOGY}\label{sec:methodology}
\subsection{fNIRS Datasets}\label{datasets}
We evaluate on three fNIRS datasets recorded under the same word-perception paradigms.

\noindent\textbf{Cao Dataset.}
Seven healthy subjects viewed $50$ words drawn from $10$ semantic categories, each
presented as a photograph of the object together with its spoken name~\cite{cao2021brain}. Every
word was shown $7$ times for $3$~s, followed by a $10$~s rest period. fNIRS
was recorded with $22$ channels over Broca's and Wernicke's areas at
$7.8$~Hz, and the released data is preprocessed.\footnote{Cao's dataset is available \href{https://www.dropbox.com/s/b7bca82ug1112km/aaai2021_cr.rar?dl=0}{here}}

\noindent\textbf{Nathan Dataset.}
Five healthy subjects perceived $6$ words from $3$ categories (vegetables, animals, and clothing). Each word was shown $20$ times for $5$~s, with a $10$~s rest. fNIRS
was recorded with $44$ channels at approximately $7.63$~Hz, with bilateral
coverage over the frontocentral, central, and temporoparietal areas. We
convert the raw intensities to optical density, apply the modified
Beer--Lambert law to recover hemoglobin concentrations, and apply a band-pass filter
between $0.01$ and $1$~Hz for consistency with the Zinszer dataset.

\noindent\textbf{Zinszer Dataset.}
This dataset~\cite{zinszer2018decoding} contains $20$ healthy
subjects who viewed $8$ words drawn from $2$ semantic categories. Every word was shown $12$ times for $3$~s, followed by a $6$ to
$9$~s rest period. fNIRS was recorded at $10$~Hz with $24$ channels in a
posterior array over the occipital lobe and $18$ channels in a lateral array
over the left temporal, parietal, and prefrontal cortex, and we use the available preprocessed data.\footnote{Zinszer's dataset is available \href{https://github.com/teammcpa/Semantic_Decoding_2017}{here}}

\subsection{Training Objectives}

The neural decoder maps fNIRS signals to vectors that approximate $d$-dimensional word embeddings of the presented stimuli. We define the decoder as a parametrized function
${f_{\theta}}: \mathbb{R}^{T\times n} \rightarrow \mathbb{R}^{d}$ that
maps an fNIRS signal $\mathbf{X} \in \mathbb{R}^{T \times n}$ of $T$ time
samples on $n$ channels, recorded while a word from a vocabulary
$\mathcal{V}$ of $V$ words is presented to the user, to a vector in
$\mathbb{R}^{f}$. We compare two training objectives: 
\begin{description}
    \item[MSE]:
\begin{equation}
    \mathcal{L}_{\text{MSE}} = \frac{1}{N} \sum_{i=1}^{N}
    \left\| \hat{\mathbf{Y}}_i - \mathbf{Y}_i \right\|_2^2,
    \label{eq:mse}
\end{equation}
\item [CLIP]: 
\begin{equation}
    \mathcal{L}_{\text{CLIP}} = -\frac{1}{N} \sum_{i=1}^{N} \log
    \left( \frac{e^{t\,s_{ii}}}{\sum_{j=1}^{N} e^{t\,s_{ij}}} \right),
    \label{eq:clip_loss}
\end{equation}
\end{description}
$N$ denotes the number of samples in a training batch, $\hat{\mathbf{Y}}_i = f_\theta(\mathbf{X}_i)$ is the embedding predicted for
the $i$-th window and $\mathbf{Y}_i$ its target, $s_{ij}$ is the cosine similarity between $\hat{\mathbf{Y}}_i$ and
$\mathbf{Y}_j$, and $t = e^{\tau}$ with $\tau$ a learnable temperature.

\subsection{Word Embeddings}

We use two pretrained word-embedding spaces as decoding targets.
\textbf{GloVe-50} provides $50$-dimensional vectors learned from global word
co-occurrence statistics~\cite{pennington2014glove}. We employ it
following~\cite{cao2021brain} for its widespread use and adequate vocabulary coverage. \textbf{T5}~\cite{raffel2020exploring}
embeddings are obtained by passing each stimulus word, in isolation and
without context, through the \texttt{t5-large} encoder and extracting the
$1024$-dimensional representation from its middle layer, as
in~\cite{d2025towards}.For \textbf{T5}, words split into multiple tokens are averaged over their token embeddings.

\subsection{Evaluation Metrics}\label{metrics}

\subsubsection{Matching Score}

Given two test trials with ground-truth embeddings $\mathbf{Y}_a, \mathbf{Y}_b$
and predictions $\hat{\mathbf{Y}}_a, \hat{\mathbf{Y}}_b$, we compare the
correct pairing, $\cos(\hat{\mathbf{Y}}_a,\mathbf{Y}_a) +
\cos(\hat{\mathbf{Y}}_b,\mathbf{Y}_b)$, against the cross-pairing,
$\cos(\hat{\mathbf{Y}}_a,\mathbf{Y}_b) + \cos(\hat{\mathbf{Y}}_b,\mathbf{Y}_a)$.
A pair is ``correct'' when the first sum is larger, and the subject-level
score is the fraction of correct pairs~\cite{cao2021brain}.

\subsubsection{Open-Vocabulary Retrieval}\label{retrieval}

Since the matching score only discriminates within a held-out pair, we also
report top-$k$ retrieval \cite{d2025towards}. For each test trial, we rank the each $w \in \mathcal{V}$" by the cosine similarity between their embedding $\mathbf{Y}_w$ and
the prediction $\hat{\mathbf{Y}}$, and count the trial as ``correct'' if the
true word is in the top $k$. We avoid top-$1$, which penalizes predicting a
synonym, and set $k$ to the number of words per category ($k=5$ for Cao,
$k=2$ for Nathan). Zinszer has only two categories of $4$ words, so we use
$k=2$ instead of $4$.

\subsubsection{Statistical Analysis}\label{stats}

Given the small number of subjects ($5$ to $20$), significance is assessed
with a one-sided Wilcoxon signed-rank test of per-subject scores against
chance. We mark $p \le 0.05$ as significant ($^{*}$) and $0.05 < p < 0.10$
as marginal ($^{\dagger}$).

\subsection{Experimental Settings}\label{setting}

\textbf{Input.} Following~\cite{cao2021brain, posso2025semantic}, we use the $1$~s HbO decoding window spanning $6.5$ to $7.5$~s after stimulus
onset, on all three datasets. Although the Zinszer and Nathan datasets also include the HbR signal, we only use HbO to match Cao's data and to
keep the input signal identical across datasets.

\noindent\textbf{Ridge-regression baseline.} Following~\cite{d2025towards}, we fit one
ridge regressor per embedding dimension, with $\alpha$ chosen by
leave-one-out cross-validation over a log grid from $10^{-2}$ to $10^{8}$.

\noindent\textbf{Bi-LSTM decoder.} We evaluate the Bi-LSTM network, which obtained the best matching scores across all experiments and subjects in~\cite{posso2025semantic}. For training, we use Adam~\cite{kingma2014adam}
under a plateau scheduler that reduces the learning rate by $10\%$ after $10$
epochs without validation improvement, with early stopping after $20$ epochs. The
architecture and remaining hyperparameters depend on the pipeline
(\S\ref{first-part}, \S\ref{second-part}).

\noindent\textbf{Training objectives.} We compare two objectives on the Bi-LSTM. The
MSE objective is regularized with an uncentered-covariance term
weighted by $\gamma$ following~\cite{posso2025semantic}. The CLIP objective uses
the same architecture and hyperparameters, so the two variants differ only
in the loss.

\subsection{Average-Trial Pipeline}\label{first-part}

We first reproduce the neural-decoding tests
from~\cite{cao2021brain, posso2025semantic} on the Cao dataset as a benchmark for the CLIP loss
and the T5 embedding, using the Bi-LSTM configuration
of~\cite{posso2025semantic}: $5$ layers, hidden size $100$, learning rate
$4.74\times10^{-3}$, weight decay $2.72\times10^{-2}$, and $\gamma=2.59$ for
MSE. The $7$ trials of each Cao word are averaged into $50$ word-vectors
per subject. The \textit{within-category} (same category) and
\textit{between-category} (different categories) tests use
leave-two-words-out evaluation: for each of the $\binom{50}{2}=1{,}225$
word pairs, we train on the remaining $48$ word-vectors and score the
held-out pair, averaging over the $100$ within and $1{,}125$ between pairs.
The \textit{leave-two-categories-out} test holds out $2$ of the $10$
categories, trains on the remaining $40$ word-vectors, and averages over the
$\binom{10}{2}=45$ pairs among the $10$ held-out words. These tests need
many categories, so we run them on Cao only; Nathan and Zinszer ($3$ and $2$
categories) are evaluated with retrieval alone (\S\ref{retrieval}).

\subsection{Single-Trial Pipeline}\label{second-part}

The average-trial pipeline is impractical for a real-world BMI, since it presupposes
repeated presentations of a word before a decoding decision can be made.
Moreover, the pairwise evaluation is a two-alternative forced choice that never asks the model to identify the word from the full vocabulary. A leave-two-words-out protocol
also scales poorly, requiring $\binom{V}{2}$ models per subject for a $V$-word
vocabulary, from $15$ for Nathan and $28$ for Zinszer to $1{,}225$ for Cao, and rising
to $\binom{350}{2}=61{,}075$ for Cao without trial averaging. We therefore re-evaluate
all models on single trials, scored by retrieval from the full vocabulary.
Each subject's trials are split word-stratified into train/validation/test
(80/10/10). For each loss, we sweep depth $L\in\{1,2,3\}$, hidden size
$H\in\{5,10,25,50,100\}$, learning rate $\eta\in\{10^{-3},10^{-2}\}$, and
embedding target $\{\text{GloVe-}50,\text{T5}\}$, tuning $\gamma$ for MSE,
and evaluate the configuration with the best validation top-$k$ accuracy
once, on the test split.

\begin{figure*}[t]
    \centering
    \includegraphics[width=\textwidth,height=4.5cm,keepaspectratio]{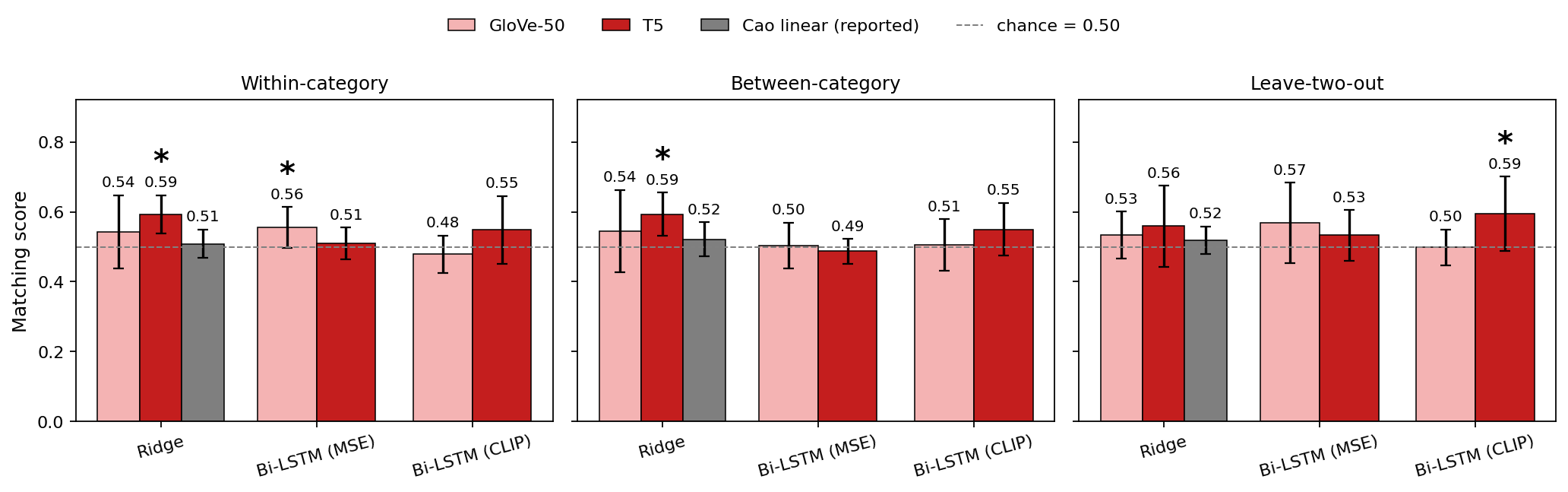}
    \caption{Matching score on Cao under the \emph{average-trial}
    pipeline (\S\ref{first-part}), averaged across subjects. 
    Significance marks are defined in \S\ref{stats}.}
    \label{fig:bars}
\end{figure*}

\section{RESULTS}\label{sec:results}

\subsection{Average-Trial Pipeline}\label{average}

As shown in Fig.~\ref{fig:bars}, replacing the GloVe-50 vectors
of~\cite{cao2021brain} with T5 improves the ridge decoder on all three
experiments, and ridge with T5 is significantly above chance on the within- and
between-category tests ($p<0.05$), as is Bi-LSTM (MSE) with GloVe-50
within-category. Bi-LSTM (MSE), however, falls to chance on the
between-category test. On the leave-two-categories-out test, where held-out
words must be decoded zero-shot, Bi-LSTM (CLIP) with T5 obtains the highest
mean ($0.59 \pm 0.11$) but is only marginally above chance ($p=0.055$). Across
all experiments, Bi-LSTM (MSE) performs better with GloVe-50, whereas Bi-LSTM
(CLIP) and ridge perform better with T5.

\begin{figure*}[t]
  \centering
  \includegraphics[width=\textwidth,height=4.5cm,keepaspectratio]{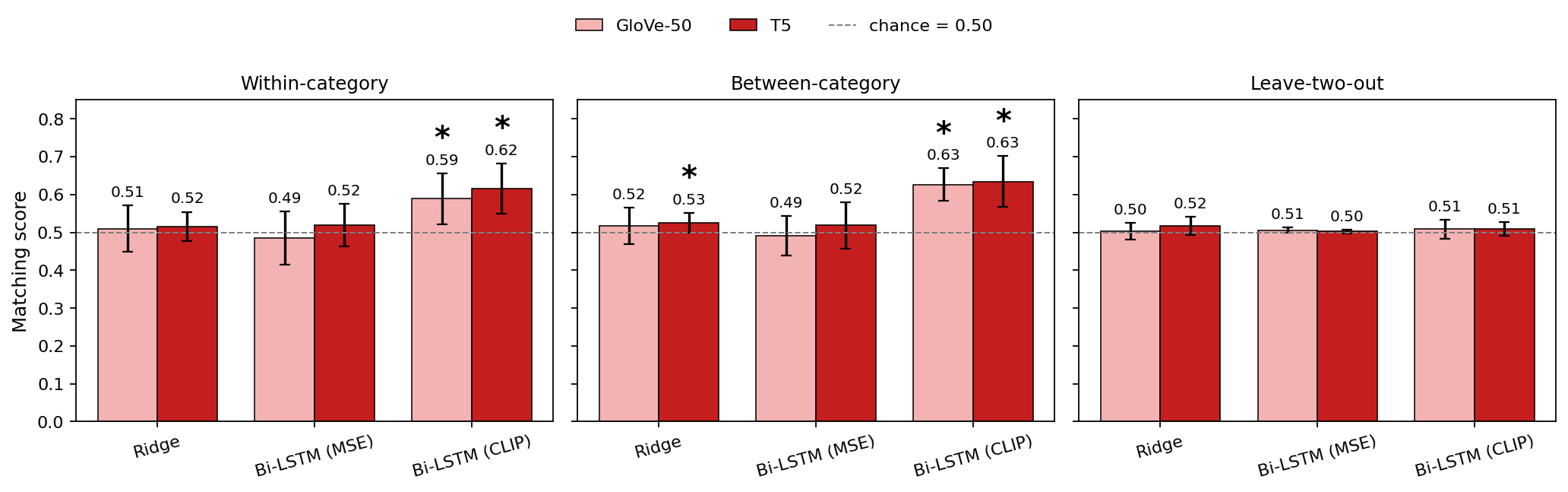}
  \caption{Matching score on the Cao dataset under the \emph{single-trial}
  pipeline (\S\ref{second-part}). Significance
  marks are defined in \S\ref{stats}.}
  \label{fig:bars-single-trial}
\end{figure*}

\subsection{Single-Trial Pipeline}\label{single}

Under the single-trial pipeline (Fig.~\ref{fig:bars-single-trial}), Bi-LSTM
(CLIP) is the only configuration significantly above chance on the within-
and between-category tests ($p<0.05$), with both embeddings, indicating that the
objective drives the gain. On the
leave-two-categories-out test every configuration falls to chance; these
scores exceed chance only under trial averaging (Fig.~\ref{fig:bars}), possibly because averaging raises the signal-to-noise ratio needed for zero-shot decoding.

\subsection{Open-Vocabulary Retrieval}\label{results-retrieval}

Tables~\ref{tab:top5}--\ref{tab:top2-zinszer} report the top-$k$ retrieval score  on all three
datasets. Bi-LSTM (CLIP) with GloVe-50 is the strongest  decoder on two of the datasets. It reaches the highest accuracy on Cao  and is the only significant configuration
on Nathan. Every significant result uses
GloVe-50, while T5 reaches at most, marginal significance. This result may be attributed to the dimensionality of the embedding, since 
T5 is larger. On Zinszer, no CLIP configuration is significant; the only one above chance is Bi-LSTM
(MSE) with GloVe-50 ($0.27 \pm 0.07$, $p=0.028$), although Bi-LSTM (CLIP)
with GloVe-50 has the highest mean ($0.28 \pm 0.11$).

At the subject level, the best single-subject score on every dataset comes
from a Bi-LSTM (CLIP) model. However, this advantage is not consistent across subjects. On
Nathan, Bi-LSTM (CLIP) with T5 has the highest mean of any configuration but also the largest between-subject variance, so it is only marginally significant ($p=0.094$).The small amount of per-subject training data is a
likely cause. After the split, each word is seen only about $5$, $16$, and
$10$ times on Cao, Nathan, and Zinszer, so each model is sensitive to its
particular split.

\begin{table}[t]
    \centering
    \caption{Top-$5$ retrieval accuracy on the Cao test set under the
    single-trial training of \S\ref{second-part}. Chance is $0.10$.
    Significance marks ($^{\boldsymbol{*}}$, $^{\boldsymbol{\dagger}}$) are
    defined in \S\ref{stats}.}
    \label{tab:top5}
    \resizebox{\columnwidth}{!}{%
    \begin{tabular}{lcc}
    \hline
    Decoder $\times$ Embedding & Top-5 mean $\pm$ std & Best subj Top-5 \\
    \hline
    Ridge $\times$ GloVe-50          & $0.10 \pm 0.02$ & $0.12$ (S3) \\
    Ridge $\times$ T5                & $0.10 \pm 0.02$ & $0.12$ (S1) \\
    Bi-LSTM (MSE) $\times$ GloVe-50  & $0.12^{\boldsymbol{*}} \pm 0.02$ & $0.13$ (S2) \\
    Bi-LSTM (MSE) $\times$ T5        & $0.08 \pm 0.02$ & $0.11$ (S7) \\
    Bi-LSTM (CLIP) $\times$ GloVe-50 & $\mathbf{0.15}^{\boldsymbol{*}} \pm 0.03$ & $\mathbf{0.19}$ (S7) \\
    Bi-LSTM (CLIP) $\times$ T5       & $0.13^{\boldsymbol{\dagger}} \pm 0.04$ & $0.19$ (S5) \\
    \hline
    \end{tabular}%
    }
\end{table}

\begin{table}[t]
    \centering
    \caption{Nathan: top-$2$ retrieval accuracy on the single-trial test set.
    Chance is $0.33$.}
    \label{tab:top2-nathan}
    \resizebox{\columnwidth}{!}{%
    \begin{tabular}{lcc}
    \hline
    Decoder $\times$ Embedding & Top-2 mean $\pm$ std & Best subj Top-2 \\
    \hline
    Ridge $\times$ GloVe-50          & $0.40 \pm 0.09$ & $0.50$ (S1) \\
    Ridge $\times$ T5                & $0.40 \pm 0.15$ & $0.67$ (S1) \\
    Bi-LSTM (MSE) $\times$ GloVe-50  & $0.38 \pm 0.13$ & $0.50$ (S4) \\
    Bi-LSTM (MSE) $\times$ T5        & $0.40^{\boldsymbol{\dagger}} \pm 0.06$ & $0.50$ (S5) \\
    Bi-LSTM (CLIP) $\times$ GloVe-50 & $\mathbf{0.49}^{\boldsymbol{*}} \pm 0.07$ & $0.56$ (S1) \\
    Bi-LSTM (CLIP) $\times$ T5       & $0.51^{\boldsymbol{\dagger}} \pm 0.17$ & $\mathbf{0.67}$ (S2) \\
    \hline
    \end{tabular}%
    }
\end{table}

\begin{table}[t]
    \centering
    \caption{Zinszer: top-$2$ retrieval accuracy on the single-trial test set.
    Chance is $0.25$.}
    \label{tab:top2-zinszer}
    \resizebox{\columnwidth}{!}{%
    \begin{tabular}{lcc}
    \hline
    Decoder $\times$ Embedding & Top-2 mean $\pm$ std & Best subj Top-2 \\
    \hline
    Ridge $\times$ GloVe-50          & $0.24 \pm 0.08$ & $0.38$ (S3) \\
    Ridge $\times$ T5                & $0.26 \pm 0.08$ & $0.38$ (S6) \\
    Bi-LSTM (MSE) $\times$ GloVe-50  & $\mathbf{0.27}^{\boldsymbol{*}} \pm 0.07$ & $0.42$ (S19) \\
    Bi-LSTM (MSE) $\times$ T5        & $0.25 \pm 0.07$ & $0.42$ (S8) \\
    Bi-LSTM (CLIP) $\times$ GloVe-50 & $0.28 \pm 0.11$ & $0.46$ (S4) \\
    Bi-LSTM (CLIP) $\times$ T5       & $0.27 \pm 0.11$ & $\mathbf{0.54}$ (S4) \\
    \hline
    \end{tabular}%
    }
\end{table}

\section{Conclusion}

This work compares the entropy-regularized-MSE and CLIP losses for reconstructing perceived words from fNIRS signals across three datasets. Bi-LSTM (CLIP) is the most robust decoder. It is the only decoder above chance on the single-trial matching score, it is significant on the retrieval task for two of the three datasets, and it matches or exceeds the best single-subject retrieval score on every dataset.
The contrastive objective suits the reconstruction task, since word decoding depends on
whether a prediction lies closer to its own target than to the alternatives,
which is what the CLIP loss optimizes directly. The two embedding targets behave differently across metrics. T5 gives the higher matching score, whereas GloVe-50 gives the more reliable retrieval accuracy despite its lower dimensionality and that T5 uses token averaging for words represented with many tokens. These results are based
on small datasets, and confirming them will require larger datasets. The CLIP objective is based 
on the InfoNCE loss~\cite{vandenoord2018infonce}, which motivates future work on other learning objectives 
that maximize mutual information between word embeddings and neural decoder mappings.

These results are a step toward accessible non-invasive BMIs that could
support communication for people who have lost the ability
to speak. This study has three main limitations. First, the datasets were collected from healthy subjects, which limits clinical translation. Second, the experimental paradigm targets language perception rather than production. Third, the stimuli are isolated words rather than natural speech.

\bibliographystyle{IEEEbib}
\bibliography{refs}

\end{document}